%  SETUP FOR LETTER

% SETUP FOR DOUBLE-SPACED PAPER

% SETUP FOR A MORE INTERESTING FONT
\def\preprintsetup{
    \paperpoint
    \doublespace
    \raggedbottom
    \toppagenumbers
    \tabskip 1em plus 1em minus 0.5em  
    \hsize=6.5truein
    \vsize=9.0in
    \rm
}
%  SETUP FOR PROPOSAL  (SINGLE SPACED)

%  NORMAL 12PT STYLE ASSIGNMENT

%  INTERESTING 12PT FONT ASSIGNMENT
\def\paperpoint{
  \font\twelvei=cmmi12
  \font\twelvesy=cmsy10 at 12pt
  \font\teni=cmmi10
  \font\tensy=cmsy10
  \font\seveni=cmmi7
  \font\sevensy=cmsy7
  \font\nrm=cmr12
  \font\rm=cmss12
  \font\it=cmti12
  \font\bf=cmbx12
  \font\sl=cmsl12
  \font\sy=cmsy10 at 12pt
  \font\twelverm=cmr12
  \font\twelvess=cmss12
  \font\tenss=cmss10
  \font\sevenss=cmr7
  \font\title=cmss17
  \textfont0= \twelvess \scriptfont0=\tenss \scriptscriptfont0=\sevenrm
  \def\rm{\fam0 \twelvess}   
  \textfont1=\twelvei  \scriptfont1=\teni \scriptscriptfont1=\seveni
  \def\mit{\fam1 } \def\oldstyle{\fam1 \twelvei}
  \textfont2=\twelvesy \scriptfont2=\tensy \scriptscriptfont2=\sevensy
  \def\doublespace{\baselineskip=24pt\parskip=0pt plus 1.5pt}
}

%  SETUP FOR FIGURE CAPTIONS
\def\begincaptions{\begingroup\parindent=0pt
   \parskip=3pt plus 1pt minus 1pt\interlinepenalty=1000\tolerance=400
   \everypar={\hangindent=0.42in\hangafter=1}}
\def\endcaptions{\endgroup\vfill\eject}
%  SETUP FOR AP.J. STYLE REFERENCES
\def\endrefs{\endgroup\vfill\eject}
\def\beginrefs{\begingroup\parindent=0pt\frenchspacing
   \parskip=1pt plus 1pt minus 1pt\interlinepenalty=1000\tolerance=400
   \everypar={\hangindent=0.42in}%\hyphenpenalty=10000
   \def\aap,##1,{{\it Astr.~Ap.,\bf ##1,}}
   \def\aaps,##1,{{\it Astr.~Ap.~Suppl.,\bf ##1,}}
   \def\aj,##1,{{\it A.~J.,\bf ##1,}}
   \def\apj,##1,{{\it Ap.~J.,\bf ##1,}}
   \def\apjl,##1,{{\it Ap.~J. (Letters),\bf ##1,}}
   \def\apjs,##1,{{\it Ap.~J.~Suppl.,\bf ##1,}}
   \def\mnras,##1,{{\it M.N.R.A.S.,\bf ##1,}}
   \def\pasp,##1,{{\it Pub.~A.S.P.,\bf ##1,}}
   \def\qjras,##1,{{\it Q.J.R.A.S.,\bf ##1,}}
   \def\josa,##1,{{\it J.~Opt.~Soc.~Am.,\bf ##1}}
   \def\araa,##1,{{\it Ann. Rev. Astr.~Ap.,\bf ##1,}}
   \def\baas,##1,{{\it Bull.~AAS,\bf ##1,}}
   \def\dap,##1,{{\it Ann. d'Ap.,\bf ##1,}}
   \def\sci,##1,{{\it Science,\bf ##1,}}
   \def\memras,##1,{{\it Mem.~R.A.S.,\bf ##1,}}
   \def\ass,##1,{{\it Astr.~Sp.~Sci.,\bf ##1,}}
   \def\nat,##1,{{\it Nature,\bf ##1,}}
   \def\jrasc,##1,{{\it J.R.A.S.C.,\bf ##1,}}
   \def\astrnach,##1,{{\it Astron.~Nach.,\bf ##1,}}
   \def\esomes,##1,{{\it ESO Messenger,\bf ##1,}}
   \def\acta,##1,{{\it Act.~Astr.,\bf ##1,}}
   \def\revmex,##1,{{\it Rev.~Mex.~A.~A.,\bf ##1,}}
   \def\aapress{{\it Astr.\ Ap.,\ }in press.}
   \def\aaletpress{{\it Astr.\ Ap.\ (Letters),\ }in press.}
   \def\aasuppress{{\it Astr.\ Ap.\ Suppl.,\ }in press.}
   \def\ajpress{{\it A.~J.,\ }in press.}
   \def\apjpress{{\it Ap.~J.,} in press.}
   \def\apjletpress{{\it Ap.~J. (Letters),} in press.}
   \def\apjsuppress{{\it Ap.~J.\ Suppl.,} in press.}
   \def\araapress{{\it Ann. Rev. A.~A.,} in press.}
   \def\baaspress{{\it Bull.~AAS,} in press.}
   \def\mnraspress{{\it M.N.R.A.S.,} in press.}
   \def\pasppress{{\it Pub.~A.S.P.,} in press.}
   \def\naturepress{{\it Nature,} in press.}
}

\def\section#1{{\bigbreak\bigskip\centerline{\bf #1}\nobreak\medskip}}
\def\subsection#1{{\medbreak\centerline{#1}\nobreak\medskip}}
\def\arcmin{{\tt '}}
    %ARCMIN WITH POINT
\def\arcsec{{\tt ''}}
\def\parcsec{{\tt ''}\mskip -7.6mu.\,}   %ARCSEC WITH POINT

   %DEGREE WITH POINT
\def\simgt{\ {\raise-.5ex\hbox{$\buildrel>\over\sim$}}\ } % < OR APPROX EQUAL TO
\def\simlt{\ {\raise-.5ex\hbox{$\buildrel<\over\sim$}}\ } % > OR APPROX EQUAL TO
% SOME USEFUL MACROS

\def\etal{{\it et~al.\ }}
\def\eg{{\it e.g.,\ }}

\def\ie{{\it i.e.,}\ }

\def\Halpha{{H$\alpha$\ }}

%  PUTS PAGENUMBERS ON TOP LIKE AP.J. PEOPLE WANT
\def\toppagenumbers{
  \nopagenumbers
  \headline={\ifnum\pageno=1\hfil\else\rm\hfil\folio\hfil\fi}
  \voffset=36pt
}
%  ALLOWS QUESTION MARKS TO BE USED AS SPACE HOLDERS WHEN NEEDED
\newdimen\digitwidth
\setbox0=\hbox{\rm0}
\digitwidth=\wd0
\tabskip 1em plus 1em minus 0.5em
%  TEX DOESN'T KNOW HOW TO HYPHENATE MY NAME
\hyphenation{Ciar-dul-lo}
\hyphenation{Kor-men-dy}

\input epsf.tex
\preprintsetup
\hsize=6.75truein
\hoffset=-0.125truein
\vsize=9.5truein
\voffset=-0.25truein
\null\vskip 0.8in
\centerline{\title PLANETARY NEBULAE AS STANDARD CANDLES. XI.}
\centerline{\title APPLICATION TO SPIRAL GALAXIES}
\bigskip
\medskip
\centerline{\title John J. Feldmeier and Robin Ciardullo$^{1,2}$}
\centerline {Department of Astronomy and Astrophysics, Penn State University}
\centerline {525 Davey Lab, University Park, Pennsylvania  16802}
\medskip
\centerline {and}
\medskip
\centerline {\title George H. Jacoby}
\centerline {Kitt Peak National Observatory, National Optical Astronomy
Observatories}
\centerline {P.O. Box 26732, Tucson, AZ 85726}
\vfill
\noindent\hangindent=0.42in
$^1$ Visiting Astronomer, Kitt Peak National Optical Astronomy, National
Optical Astronomical Observatories, operated by the Association of
Universities for Research in Astronomy, Inc., under cooperative agreement
with the National Science Foundation.

\noindent\hangindent=0.42in
$^2$NSF Young Investigator
\eject

\section{Abstract}

We report the results of an [O~III] $\lambda 5007$ survey for planetary
nebulae (PN) in three spiral galaxies: M101 (NGC~5457), M51 (NGC~5194/5195) 
and M96 (NGC~3368).  By comparing on-band/off-band [O~III] $\lambda 5007$ 
images with images taken in \Halpha and broadband $R$, we identify 
65, 64 and 74~PN candidates in each galaxy, respectively.  From 
these data, an adopted M31 distance of 770~kpc, and the empirical planetary 
nebula luminosity function (PNLF), we derive distances to M101, M51, and M96 
of $7.7 \pm  0.5$, $8.4 \pm 0.6$, and $9.6 \pm 0.6$~Mpc. These 
observations demonstrate that the PNLF technique can be successfully applied 
to late-type galaxies, and provide an important overlap between the 
Population~I and Population~II distance scales.  We also 
discuss some special problems associated with using the PNLF in spiral
galaxies, including the effects of dust and the possible presence of
[O~III] bright supernova remnants.

\vskip 24pt
\noindent{\it Subject Headings:  }galaxies: distances and redshifts --- 
distance scale --- galaxies: individual (M101, NGC 5457, M51, NGC 5194,
M96, NGC 3368) --- planetary nebulae: general --- supernova remnants

\vfill\eject 

\section{1. Introduction}

The planetary nebula luminosity function (PNLF) is a distance indicator
for galaxies potentially as far away as $\approx 25$ Mpc.  It has 
proved to be remarkably successful for many galaxies, including galaxies
in the Virgo cluster (Jacoby, Ciardullo, \& Ford 1990a), the Fornax
cluster (McMillan, Ciardullo, \& Jacoby 1993), and the Magellanic Clouds
(Jacoby, Walker, \& Ciardullo 1990b).  For a complete review, see Jacoby 
\etal (1992).

The PNLF uses the observational result, supported by theory (Jacoby 1989; 
Han, Podsiadlowski, \& Eggleton 1994; M\'endez \etal 1993; Stanghellini 1995),
that the luminosity function of planetary nebulae (PN), in the light of [O~III] 
$\lambda 5007$, has the form
$$ N(M) \propto e^{0.307M} \, [1 - e^{3(M^{*}-M)}] \eqno(1) $$
(Ciardullo \etal 1989a).  Assuming that this function is correct, we can 
compare the PNLF in a distant galaxy to a PNLF in a nearby galaxy 
whose distance is already known from some other method.  
The calibration for the PNLF is currently provided by M31, whose distance is 
well known from Cepheid variables (Freedman \& Madore 1990).  This calibration
has been validated by checks in M81 (Jacoby \etal 1989), the LMC 
(Jacoby, Walker, \& Ciardullo 1990b), and NGC~300 (Soffner \etal 1996).  

Until now, the PNLF technique has 
been used primarily in elliptical and S0 galaxies, where the problems presented
by internal extinction and interloping H~II regions are minimal.  However, 
since planetary nebulae are not associated with any one stellar population, 
PN measurements in spirals should be possible.  Specifically, it should be 
possible to discriminate planetary nebulae from H~II regions on the basis of 
their [O~III]/$H\alpha$ line ratio.  Bright planetary nebulae are high 
excitation objects, hence $I(\lambda 5007)/I(H\alpha) > 1$; the opposite is 
true for H~II regions.  

There are a number of reasons to attempt to apply the PNLF technique to spiral
galaxies.  First, observing the PNLF in more luminous galaxies 
allows us to check for internal consistency of the method.  
Critics have proposed that the maximum
magnitude of the PNLF, $M^*$, increases with the absolute magnitude of
the galaxy (Bottinelli \etal 1991).  Clearly, observing bright galaxies
along with dimmer galaxies in a cluster would test this
hypothesis.  This has already been done in the Fornax cluster
(McMillan, Ciardullo, \& Jacoby 1993), and in the Virgo cluster 
(Jacoby, Ciardullo, \& Ford 1990a), but doing more high luminosity galaxies, 
such as bright spirals, would constrain the invariance of $M^*$ even further.  

Secondly, since the PNLF has been primarily used in old stellar
populations, it is possible that $M^*$ depends on population age.  
Han, Podsiadlowski, \& Eggleton (1994), M\'endez \etal (1993), Stanghellini 
(1995), and Soffner \etal (1996) have all proposed that $M^{*}$ may be 
brighter in younger stellar populations.
The way to test this hypothesis is to observe galaxies with 
ongoing star formation, such as late-type spirals.

Finally, the most important test of any distance method is its comparison with 
other methods.  A fundamental weakness of the current extragalactic 
distance scale is the small amount of overlap and cross-checks between some 
of the techniques.  In fact, there are actually two distinct distance scales:
a Population~II scale, defined by the planetary nebula luminosity function
(PNLF), surface brightness fluctuations (SBF), the globular cluster
luminosity function (GCLF), and the elliptical galaxy fundamental plane ($D_n$-$\sigma$) relations, and a Population~I scale, involving Cepheid 
variables, supernovae, and the Tully-Fisher relation.  Remarkably, there 
are only five galaxies common to both systems: M31, the Large Magellanic 
Cloud, NGC 5253, M81, and NGC 300.  Despite all the checks and comparisons,
the calibration of the Population~II distance scale rests solely 
on these five galaxies.  All other calibrations are indirect
(\ie they use different galaxies within a common cluster), and are thus
susceptible to systematic errors due to galaxy segregation.

The best method of linking the two distance scales is through
the planetary nebula luminosity function.  Unlike Cepheids, which are
exclusively Population~I objects, or the SBF method, which can only be
applied in relatively red galaxies with smooth isophotal profiles, 
PNLF measurements can
(in theory) be performed in any galaxy, as long as there are enough 
bright PN to populate the bright end of the luminosity function.  
Thus, the PNLF has the potential to be {\it both\/} a 
Population I and II distance indicator, assuming there are no systematic
biases with galaxy type.  The purpose of this paper is to test
whether this is true.  

\section{2. Observations and Data Reduction}

The observations were obtained on 1995 April 5-8, and 11 using the prime 
focus of the Kitt Peak 4-m telescope and the T2KB 2048 x 2048 Tektronix CCD,
which has a pixel scale of $0\parcsec 47$ per pixel and a field-of-view of
$16' \times 16'$.  We obtained exposures for three galaxies, M101 (NGC~5457),
M51 (NGC~5194 and NGC~5195), and M96 (NGC~3368), through a 30~\AA\ wide 
[O~III] $\lambda 5007$ filter centered on the galaxies' systemic velocities.
Corresponding images were then taken through a 275~\AA\ wide 
off-band filter (central wavelength $\sim 5300$~\AA), a 75~\AA\ wide \Halpha 
filter, and a broadband $R$ filter.  The seeing throughout the observations
was always better than $1\parcsec 2$.  A log of these observations is 
given in Table~1.  The data were bias-subtracted and flat-fielded at the
telescope, using the IRAF reduction system.

Our survey technique was as described in previous papers
(\eg Jacoby \etal 1989; Ciardullo, Jacoby, \& Ford 1989b).
We identified PN candidates by ``blinking'' the sum of the on-band
[O~III] images for each galaxy against a corresponding off-band sum.  
In order to discriminate planetary nebulae from other emission line 
sources, we used the following criteria:  1) PN candidates had to have
a point-spread-function (PSF) consistent with that of a point source;
2) PN candidates had to be present on the [O~III] $\lambda 5007$ image,
but invisible on the off-band image; and 3) PN candidates
had to be invisible in $R$ and invisible in H$\alpha$.
In addition, to further reduce the possibility of contamination 
from H~II regions, we did not attempt to find planetary nebulae within the
obvious star-forming regions of the galaxies (\ie the spiral arms); instead,
we concentrated our search in the galactic halos and inter-arm regions.  

By applying the above criteria, we severely reduced the possibility of an 
H~II region accidentally being included in our list of planetary candidates.
At a distance of $\sim$ 10 Mpc, one arcsecond is equivalent to 
$\sim$ 48 parsecs, thus we resolved the majority of H~II regions out of our
sample.  Similarly, by requiring that PN candidates be invisible on the 
off-band and $R$-band frames, we excluded those emission-line 
regions that have bright, exciting OB-stars.  Our most valuable condition, 
however, was the [O~III] $\lambda 5007$ to \Halpha emission line
ratio.  Unlike bright planetary nebulae, H~II regions are usually 
low excitation 
objects with $I(\lambda 5007)/I({\rm H}\alpha) < 1$.  However, to be
considered a planetary nebula candidate, an object had to have an
[O~III] $\lambda 5007$ to H$\alpha$ ratio greater than 1.6.  (For the
faintest PN in the sample, $I(\lambda 5007)/I({\rm H}\alpha)$ had to be
$\sim 1.6$, which was the ratio of the detection limits of the two images;
for brighter PN, the minimum line ratio was greater, since we required
that all PN candidates be invisible in H$\alpha$.)  In the Shaver \etal (1983)
sample of Galactic H~II regions, only 11\% of the objects have an [O~III] to 
H$\alpha$ ratio greater than 1.6, and only one out of 39 has
$I(\lambda 5007)/I({\rm H}\alpha) > 3$.  Conversely, the [O~III]
$\lambda 5007$ PNLFs defined in other galaxies (\eg M31: Ciardullo \etal 1989a; M81: Jacoby \etal 1989; NGC~891: Ciardullo, Jacoby, \& Harris 1991; NGC~4565 
and 4278: Jacoby, Ciardullo, \& Harris 1996; M87: Ciardullo, Jacoby, \& 
Feldmeier 1997) include only those objects with 
$I(\lambda 5007)/I({\rm H}\alpha) \simgt 2.0$.  Thus, by excluding those
objects that are \Halpha bright, we removed most of the
H~II regions from the sample while leaving the bright end of the
[O~III] PNLF intact.

With all the above constraints, only compact, inter-arm H~II regions that
have high nebula excitation and faint central OB associations will have been
mistaken for planetary nebulae.  However, even these objects will
not significantly affect our PNLF{}.  Because our threshold 
$I(\lambda 5007)/I({\rm H}\alpha)$ ratio is more stringent for brighter
objects, any H~II regions that do make it into our sample will be 
faint in [O~III] and 
fall below our photometric completeness limit (described below).  We are 
therefore confident that essentially all the objects in our sample
are, indeed planetary nebulae, and not H~II regions.  Furthermore, as we
shall see below, the agreement between the observed shapes of the PNLF's,
and the agreement between the PNLF and Cepheid distances
provides a validity check on our H~II region exclusion criteria.  Had the
PN sample been seriously contaminated by bright H~II regions the PNLF fits
to the empirical law would have been poor and our distances
would have been systematically and unpredictably too small.  

In total, we identified 65 candidate planetaries in M101, 64 in M51, 
and 74 in M96.  
Sample images of candidate planetary nebulae in the light of
[O~III] $\lambda 5007$, H$\alpha$, and offband [O~III] are given in 
Figure~3 of Feldmeier, Ciardullo, \& Jacoby (1996).   

The PN candidates were measured photometrically using the IRAF version
of DAOPHOT (Stetson 1987), and flux calibrated using Stone (1977) standard 
stars and the procedures outlined by Jacoby, Quigley, \& Africano (1987).  
The resulting monochromatic fluxes were then converted to $m_{5007}$ 
magnitudes using
$$ m_{5007} = -2.5 \log F_{5007} - 13.74 \eqno(2) $$

Astrometry was performed using the positions of stars in the {\sl Hubble 
Space Telescope\/} Guide Star Catalog (GSC).  In most cases, the GSC 
stars were saturated, requiring us to use a two-step process.
First, we used digitized images of the Palomar Sky Survey to measure the
positions of unsaturated stars on our CCD frames relative to stars in the
GSC catalog.  We then found the positions of our PN candidates relative
to these secondary standards.  Based on the residuals of our fits, we 
estimate that these positions are accurate to $ 0\parcsec 68$ in M101, 
$ 0\parcsec 49$ in M51, and $ 0\parcsec 44$ in M96.

\section{3. Fitting the PNLFs and Obtaining Distances}

Before we can fit our observed luminosity functions to equation (1),
we must estimate our photometric completeness level.
Because our PN survey was conducted primarily in the low surface-brightness
regions of each galaxy, our ability to detect PN was not a strong function of
galactic position.  Thus, we used the results of Jacoby \etal (1989) and Hui 
\etal (1993) to estimate our limiting magnitude for completeness as the place 
where the PNLF (which should be exponentially increasing) begins to turn down.
To confirm this number, we randomly added artificial stars to our summed 
on-band images, and reblinked the frames to recover as many of the simulated
objects as possible.  As expected, the fraction of our recoveries was 
independent of PN magnitude down to about the completeness level, except in
the spiral arms and innermost regions of the galaxies.

Unfortunately, due to the complex and variable backgrounds presented by our 
target galaxies, the detectability of each planetary is different, and 
this could possibly change the shape of the observed luminosity function.
To deal with this problem, we created a statistical sub-sample of PN
in each galaxy, for which the effects of the sky background were minimal.
To form this sample, we began by noting the 
median sky background associated with each PN measurement.  After
excluding those few objects superposed on bright regions of the galaxy,
we picked the worst (most uncertain) background remaining in the sample, and
computed the signal-to-noise each PN would have, if it were projected on 
that background.  Only those objects which would have been detected with a 
signal-to-noise greater than 10 (cf.~Ciardullo \etal 1987; Hui \etal 1993)
on that difficult background were included in our analysis.  This left
27 planetaries in M101, 42~PN in M51, and 33~PN in M96 available for
further analysis.  Lists of all the planetaries found, their positions
and magnitudes, and whether they are members of the statistical sample, 
are given in Tables 2, 3 and 4.  Those PN that are part of the statistical
samples are marked with an ``S.''

The PNLF distance to each galaxy and the formal uncertainty in each case
were calculated by convolving the empirical function given in equation (1)
with a photometric error vs.~magnitude relation derived from the output
of DAOPHOT, and fitting the resultant curve to the complete sample of PN,
via the method of maximum likelihood (Ciardullo \etal 1989a).  In order
to place our distances on the same system as previous PNLF distance
determinations, we adopt $M^* = -4.48$, based on an M31 distance of 710~kpc
(Welch \etal 1986), a foreground M31 reddening of $E(B$$-$$V) = 0.11$ 
(McClure \& Racine 1969), and a Seaton (1979) reddening curve.  With more 
recent values for M31's distance (770~kpc; Freedman \& Madore 1990) and 
reddening ($E(B$$-$$V) = 0.08$; Burstein \& Heiles 1984), the distances 
reported here would increase by $\sim 3$\%, or $+ 0.06$ mag in distance
modulus.  All our galaxy distances have assumed the foreground extinction
values of Burstein \& Heiles (1984). 

The formal errors associated with our maximum-likelihood PNLF fits 
are +0.04/$-0.07$~mag for M101, +0.06/$-0.10$~mag for M51 and
+0.05/$-0.07$~mag for M96.  To compute the total error budget,
these errors must be combined in quadrature with those associated 
with photometric zero point (0.02~mag), the 
filter response curve (0.04~mag), and the Galactic foreground extinction 
(0.05~mag; from Burstein \& Heiles 1984).  In addition, two systematic errors, 
which affect all PNLF measurements the same way, arise from the uncertain 
definition of the empirical PNLF (0.05~mag), and, of course, the distance to 
the calibration galaxy M31 (0.10~mag).  The observed planetary nebula
luminosity functions for the statistically complete samples of objects
are displayed in Figure~1, along with the best fits for
the empirical law.  The luminosity functions are qualitatively similar
to all the PNLF's seen in elliptical galaxies. 

\section{4. Results and Discussion}

\subsection{4.1. M101}

The PNLF distance to M101 was reported in an earlier paper
(Feldmeier, Ciardullo, \& Jacoby 1996), so here we focus on additional results.
The best-fitting distance for M101 (assuming no Galactic
extinction; Burstein \& Heiles 1984) is $(m$$-$$M)_0 = 29.36 \pm 0.15$,
including all errors.  With the newer M31 distance,
this number increases to ($m$$-$$M)_0 = 29.42 \pm 0.15$.  This distance
assumes that the PNLF has not been affected by M101's internal extinction.
If we adopt the mean measured extinction to Cepheids observed by
the {\sl Hubble Space Telescope\/} ($E(B$$-$$V)=0.03$; Kelson \etal 1996), the 
distance modulus drops to within 1\% of the {\sl HST\/} Cepheid 
result ($(m$$-$$M)_0 = 29.34 \pm 0.16$). The issue of internal 
extinction is discussed in detail in \S 5.  

Given M101's highly irregular background, it is extremely difficult to 
determine $\alpha_{2.5}$, the number of planetaries within 2.5 magnitudes of 
$M^*$ normalized to the galaxy's bolometric luminosity.  We can, however, 
estimate a lower limit to the number of planetaries expected within our 
detection threshold, and compare that to the number of PN actually found, 
assuming some value of $\alpha_{2.5}$.  To do this, we first calculate the 
approximate $B$ luminosity of M101, using the galaxy's total $B$ magnitude of 
8.31 (de Vaucouleurs \etal 1991; Buta \etal 1995), and our distance modulus 
of 29.42.  If we assume the difference between M101's $B$ and bolometric 
magnitudes is small, these numbers imply a total bolometric luminosity for 
the galaxy of $2 \times 10^{10} L_{\odot}$.  If we then adopt a mean 
population age for the PN progenitors of $2 \times 10^8$~years, then the 
models of Renzini \& Buzzoni (1986) imply a specific evolutionary flux of 
$\sim 1 \times 10^{-11}$~stars yr$^{-1} L_{\odot}^{-1}$.  If it takes 
$\sim 25,000$ years for PN to fade 8~mag below $M^*$, equation (1) then 
implies that we should see $\sim 125$~PN in the brightest $\sim 1$~mag of 
the PNLF\null.  We note that this estimate is extremely rough, as it depends 
on the uncertain bolometric luminosity of the galaxy, the assumption that all
stars eventually become [O~III] bright PN, and a young age for the PN 
progenitors.  Nonetheless, our observations of 
65 PN suggest that we are not sampling all the M101 PN brighter than our 
detection limit. This is to be expected, since the [O~III] emission from H~II 
regions interferes with finding planetaries in the spiral arms, and the high 
surface brightness prevents us from finding planetaries near the galaxy's 
nucleus.  Moreover, it is possible that not all of M101's stars evolve through
the [O~III] bright PN phase. Ciardullo (1995) has found that the PN production
rate in metal-rich, UV-bright elliptical galaxies can be a factor of 
$\sim 4$ less than predicted.  Although no star-forming galaxy has ever been
observed to be PN deficient, M101 is the first metal-rich spiral to be
surveyed.  Thus, although we are missing planetary nebulae, it is
impossible to quantify exactly what our sampling rate is.  However, these 
lack of detections should not affect the derived PNLF distance unless there 
are systematic differences between the detected and undetected PN.

\subsection{\it 4.1.1. Comparisons of M101 Distances}

As there are so many distance estimates to this Sc galaxy, and given 
its well known importance for determining the Hubble constant, it is
appropriate to compare our distance to M101 with distances derived from 
other techniques.  We first compare our results to the Cepheid variables,
which are thought to be the most reliable extragalactic distance indicator.
From their inability to detect Cepheid variables photographically,
Sandage \& Tammann (1974a) first claimed an M101 distance modulus of 
$(m$$-$$M)_0 > 29.0.$  Much later, Cook, Aaronson, \& Illingworth (1986) 
used a CCD to detect 2 Cepheids in the R-band, and derived a preliminary 
distance modulus of $\sim 29.2$.  This was supplemented by observations
of two additional Cepheids, and five Mira variable stars observed by Alves 
\& Cook (1995), who found $(m$$-$$M)_0 = 29.08 \pm 0.13$, a value somewhat 
smaller than Cohen's (1993) Cepheid estimate of $(m$$-$$M)_0 = 29.4 \pm 0.15$.
The most reliable Cepheid distance measurement, however, comes from
Kelson \etal (1996) as part of the {\sl Hubble Space Telescope\/} Distance
Scale Key Project.  Kelson \etal detected 29 Cepheids in the outer part of 
M101, and derived a distance modulus of $(m$$-$$M)_0 = 29.34 \pm 0.17$, with 
a mean Cepheid reddening of $E(B$$-$$V) = 0.03$.  This distance measurement 
is in excellent agreement with our value.  

We now briefly compare our PNLF distance to some other methods.  
Using the expanding photosphere method (EPM) on SN~1970G, Schmidt, 
Kirshner, \& Eastman (1992) found a distance modulus to M101 of  
$(m$$-$$M)_0 = 29.34^{+0.28}_{-0.49}$.  Though the errors are
large, the agreement with both the PNLF method and the HST Cepheids
is encouraging, and suggests that the EPM method is 
fundamentally sound.  From a mean BRI Tully-Fisher relation,
Pierce (1994) reported a rough distance modulus of 
$(m$$-$$M)_0 = 29.2 \pm 0.5$, also in good agreement with our
result.  Finally, M101 has been used as a testbed for using the 
brightest stars as distance indicators
(Sandage 1983; Humphreys \& Aaronson 1987).  Though the errors in this 
method are thought to be large (Rozanski \& Rowan-Robinson 1994),
the assertion by Humphreys \etal (1986) that the distance modulus 
of M101 must be less than 28.4 because of the limits imposed 
by the intrinsic luminosities of M supergiants, seems to be incorrect.
The distances to M101 given in the literature are listed
in Table~5.  For some comments on the earlier results, see de
Vaucouleurs (1993).

The extremely good agreement between M101's {\sl HST\/} Cepheid distance and
our PNLF distance puts a strong limit on the systematic effect proposed by 
Bottinelli \etal (1991).  Bottinelli \etal hypothesized that the true
shape of the planetary nebula luminosity function has a power-law slope
at the bright end, with no physical cutoff.  Consequently, their model
predicts that all PNLF distances have a systematic error, which is 
correlated with the absolute luminosity of the host galaxy.  
Bottinelli \etal parameterized this effect (their equation~3), as
$$ \mu^{'}  = (0.4/\alpha^{'}) (M_{g} - M^{0}_{g})   + \mu_{true} \eqno(3) $$
where $M_g$ and $M^0_g$ are the absolute magnitudes of 
an unknown galaxy and of a galaxy where the value of
$M^*$ is known precisely, and  $\alpha^{'} = 1.6$.
We can test this formula by calculating the predicted shift due to
this supposed effect, and comparing it to our actual results.  
We follow the methodology of Bottinelli \etal (1991), and substitute
$M_B$ magnitudes for $M_g$ and $M^0_g$, even though this
violates the equation above.  The calibrator galaxy for the PNLF, 
the bulge of M31, has $B_T = 4.92$ (de Vaucouleurs 1958) with $A_B = 0.32$
(Burstein \& Heiles 1984), and we assume a distance modulus to M31 of 
$(m$$-$$M)_0 = 24.42 \pm 0.12$ (Freedman \& Madore 1990).  If we adopt the 
Cepheid distance modulus to M101 of $(m$$-$$M)_0 = 29.34 \pm 0.16$ 
(Kelson \etal 1996), and use $B^0_T = 8.31$ (de Vaucouleurs \etal 1991; 
Buta \etal 1995), then we find the expected error in our M101 distance is:
$$ \mu^{'}  =  -0.31 \pm 0.05   + \mu_{true} \eqno(4) $$
Our derived distance of $(m$$-$$M)_0 = 29.42 \pm 0.15$,
disagrees with this estimate by $0.39 \pm 0.16$ magnitudes.  
Even more significantly, if the M101 Cepheid distance is precisely correct,
our PNLF distance is a slight overestimate.  If the Bottinelli \etal 
hypothesis were true, we should underestimate the distance to M101 compared 
to the Cepheid distance.  From the results above, the predicted shift of
Bottinelli \etal is excluded with greater than 98\% confidence.  This, 
along with other studies in the Virgo and Fornax clusters, invalidates
the Bottinelli \etal premise.   

\subsection{4.2. M51}

The M51 system is a galaxy pair consisting of M51 (NGC~5194), an 
SBbc(s) II-III spiral, and
NGC~5195, a satellite companion of type SB0~pec (Sandage \& Tammann 1987). 
The main body of M51 has a higher surface brightness than M101 and
a smaller inter-arm spacing, so planetary detections in these
areas are difficult.  However, as Figure~2 shows, a substantial
planetary nebula population exists in M51's halo.  Interestingly, the
largest concentration of planetary nebulae is to the west of NGC~5195.
Deep images of the M51 system show an increase in surface brightness in 
this region (for an example, see Burkhead 1978),
but it is unknown whether the planetaries are related to NGC~5195, or
are part of a halo population of M51.  This pattern is most
likely due to the tidal interaction of the two galaxies, but in 
theoretical studies (Toomre \& Toomre 1972; Howard \& Byrd 1990)
this area has not been mentioned.
Radial velocity measurements of these planetaries should 
provide an interesting dynamical constraint to the system.

The PNLF for this galaxy is plotted in Figure~1.   
The derived best-fitting distance to M51 (assuming no Galactic extinction;
Burstein \& Heiles 1984) is $(m$$-$$M)_0 = 29.56 \pm 0.15$, including
all errors.  With the improved M31 distance, our value increases  to
$(m$$-$$M)_0 = 29.62 \pm 0.15$.
This result also assumes that internal extinction in the galaxy is
insignificant.  Since almost all of the PN are outside
of the visible optical disk, this is likely to be a valid assumption
(Giovanelli \etal 1994).

\subsection{\it 4.2.1. Comparisons of M51 Distances}

Despite the well known nature of the M51 galaxy system and its importance, 
few distance determinations exist in the literature.  
In the past, most distances to this galaxy have been based on 
group membership, or an assumed value of the Hubble constant; these methods
generally placed M51 near, or slightly more distant than M101 
(Sandage \& Tammann 1974a; Holmberg 1964).  The only two direct measurements 
of M51's distance prior to 1994 disagreed dramatically.  The first, by Sandage 
\& Tammann (1974b), placed M51 at $(m$$-$$M)_0 = 29.91$ based on the sizes of 
its H~II regions; the second by Georgiev \etal (1990), put the galaxy at $(m$$-$$M)_0 = 29.2 \pm 0.20$ via the sizes of its young stellar associations.
Fortunately, in the last year three new distance estimates have been made.  
Baron \etal 
(1996) used the spectral-fitting expanding atmosphere method (SEAM) on the
type Ic SN~1994I, to
derive a preliminary distance modulus of $(m$$-$$M)_0 =  28.9 + 5 \log \, 
(t_r/9$~days$) \pm 0.5 \pm 0.7$, where $t_r$ is the rise time of the bolometric
light curve, the first set of uncertainties is due to the model, and
the second set of uncertainties includes the effects of extinction.
(They assume $E(B$$-$$V)_{SN} = 0.45$, and their best value for $t_{r}$ is 9 days.)  This distance is thought to be preliminary because of the sensitivity 
of the models to abundance stratification (Baron 1996).  Similarly,
Iwamoto \etal (1994) derived a distance modulus to SN 1994I of 
$(m$$-$$M)_0 = 29.2 \pm 0.3$
from their theoretical calculations of exploding C + O stars.  Unfortunately,
both these distance determinations suffer from the fact that SN~1994I 
occurred in a dust lane close to the center of the galaxy, and thus the
value of the extinction is highly uncertain (Richmond \etal 1996).

Perhaps the most reliable distance to M51 comes from surface brightness 
fluctuations found by Tonry (1996).  The preliminary distance to NGC~5195
based on this method is $(m$$-$$M)_0 = 29.59 \pm 0.15$,
in excellent agreement with our value.  
The distances to M51 are given in Table~6.
 
\subsection{4.3. M96}

M96 is an Sab(s)II galaxy in the Leo~I group of galaxies.  Leo~I
has been recognized as an important stepping stone to the
Hubble constant as far back as Humason, Mayall, \& Sandage (1956), and
its brightest spiral, M96, has gained new attention due to the recent
{\sl Hubble Space Telescope\/} Cepheid measurements by Tanvir \etal (1995).
In Figure~3, we plot the positions of M96's planetary nebulae candidates.
Once again, the planetaries are detected at large galactocentric radii, far
away from the high surface brightness regions of the galaxy. This is the most 
distant of the galaxies in this survey, and the data were taken under the poorest seeing conditions: only $\sim 0.8$~mag of the PNLF is sampled.  
Nevertheless, we derive a best-fitting distance to M96 (assuming a Galactic 
extinction of 0.015 mag; Burstein \& Heiles 1984) of $(m$$-$$M)_0 = 29.85 \pm 
0.15$, including all errors.  With the improved distance to M31, this distance 
increases to $(m$$-$$M)_0 = 29.91 \pm 0.15$.  As for M51, we make no 
correction for internal extinction as the amount of dust at large radii 
is likely to be small (Giovanelli \etal 1994).

\subsection{\it 4.3.1. Comparisons of M96 Distances}

The Leo I group is a close-by, well-mixed galaxy group, containing 
two giant ellipticals (NGC~3377 and 3379), two SB0's (NGC~3384 and 3412), 
an Sab (NGC~3368), and an SBb (NGC~3351) galaxy.  It is felt from many studies
(Turner \& Gott 1976; Huchra \& Geller 1982; Geller \& Huchra 1983; Vennik
1984; Tully 1988; Schneider 1989; Schneider \etal 1989) 
that these galaxies are at the same approximate distance.  This allows us 
to compare our M96 distance to Leo~I group distances founds from other 
methods.  Historically, these distances have ranged from as high as 
$(m$$-$$M)_0 = 31.18$ (Visvanathan 1979), to as low as $(m$$-$$M)_0 = 29.84$ 
(Tonry 1991).  We now focus on several selected distance indicators to the 
group.  

The Tully-Fisher relation has been partially calibrated on galaxies
near the Leo~I group.  Specifically, the nearby 
Leo triplet, which consists of 
NGC~3623, 3627, and 3628, has been used by Aaronson \& Mould (1983)
as a infrared Tully-Fisher calibrator.  Their group
distance modulus of $(m$$-$$M)_0 = 29.84 \pm 0.25$, with an additional
$\sim 0.2$~mag systematic uncertainty, is in good agreement with our
value, though it is problematic whether this triplet is at the same 
distance as the Leo I group.

The globular cluster luminosity function (GCLF) distance
indicator has also been used on galaxies in the Leo~I group.  By summing
the globular clusters found in NGC 3377 and NGC 3379, and assuming a Milky
Way GCLF peak at $M^{0}_{B} = -6.84 \pm 0.17$, Harris (1990) estimated the 
group's distance to be $(m$$-$$M)_0 = 30.19 \pm 0.43$.  This result
is also in approximate agreement with our M96 result, especially considering
the method's large error bars.

Another standard candle that has recently been used in the Leo~I Group is
the tip of the red giant branch.  By identifying the abrupt discontinuity
in the luminosity function of resolved stars in NGC~3379, Sakai \etal (1996) 
have estimated a distance to the group of $(m$$-$$M)_0 = 30.20 \pm 0.14$.
Again, this is consistent with the results from planetary nebulae.

A technique that has been compared frequently to the PNLF is the
surface brightness fluctuations (SBF) method.  This method,
pioneered by Tonry \& Schneider (1988), has been used on a large
number of early-type galaxies, including NGC~3377, 3379, and
3384 (Tonry \etal 1990; Tonry 1991; Ciardullo, Jacoby, \& Tonry 1993).  
The best-fitting SBF distance 
moduli for these galaxies are $(m$$-$$M)_0 = 29.94 \pm 0.08$,
$(m$$-$$M)_0 = 29.87 \pm 0.07$, and $(m$$-$$M)_0 = 29.88 \pm 0.12$
respectively, in excellent agreement with our distance to M96.  A discussion 
of PNLF versus SBF distances is given in Ciardullo, Jacoby, \& 
Tonry (1993).

A critical test for the PNLF method is the comparison of our M96 distance
with the PNLF measurements to the other galaxies of the group.  Our
M96 distance modulus of $(m$$-$$M)_0 = 29.91 \pm 0.15$ is, within the errors,
identical to that previously measured to the three early-type galaxies
NGC~3377, 3379, and 3384 (distance moduli of $30.13 \pm 0.17$,
$30.02 \pm 0.16$, and $30.09 \pm 0.15$, respectively).  This agreement
further reinforces the conclusion that PNLF distances have little or no 
dependence on galaxy Hubble type (cf.~Ciardullo, Jacoby, \& Harris 1991; 
Jacoby, Ciardullo, \& Harris 1996).

Finally, we look at the {\sl Hubble Space Telescope\/} Cepheid distances 
to the group.  There are two Leo~I galaxies with {\sl HST\/} Cepheid 
measurements: M96 and M95 (NGC~3351).  Unfortunately, these two measurements 
yield conflicting results.  From the light curves of 45 Cepheids found
in M95, Graham \etal (1996) estimate a group distance modulus of 
$(m$$-$$M)_0 = 30.01 \pm 0.19$, in good agreement to our value.  However, 
the initial results derived from the analysis of seven Cepheids in M96 (Tanvir
\etal 1995) imply a distance modulus of $(m$$-$$M)_0 = 30.32 \pm 0.16$, 
$1.9 \sigma$ larger than the PNLF distance.  This PNLF/Cepheid distance 
discrepancy is the largest to date, and it occurs in the most distant galaxy 
with both a PNLF and Cepheid distance determination.   If the M96 Cepheid 
distance is correct, it may be the first evidence for a change of $M^*$ 
with stellar population.  However, it is difficult to understand why 
such an effect is not seen in M101 or M51. 

Alternatively, Graham \etal claim that the $V$-band Cepheid period-luminosity
relations found for M95 and M96 are consistent, and that only the $I$-band 
relations are different.  Thus, the only difference between the two numbers
is the derived values of the absorption corrections.  Madore \& Freedman (1991)
state that in order to obtain both an accurate distance and reddening estimate
to a galaxy, two or three dozen Cepheids are needed.  Since only seven 
Cepheids are used to estimate the Cepheid distance to M96, (and only 3 
$I$-band measurements were obtained), it is possible that the Cepheid 
distance to this galaxy has been overestimated.  In Table 7, we
list the distances to Leo I group galaxies given in the literature.
 
\section{5. The Effect of Dust}

In all the above distance determinations, we ignored the presence
of dust within the target galaxy.  Several lines of reasoning suggest
this is reasonable.  First, our method for identifying a 
statistically complete sample of planetary nebulae selects against 
objects superposed near regions of strong extinction.  By excluding 
objects that have highly variable backgrounds, we select against 
planetary nebulae that are close to patches of dust.

Second, there is evidence from other studies that the disks of spiral 
galaxies are not opaque, though there is considerable debate on this point
(for opposing views, see Valentijn 1990 and Disney, Davies, \& Phillips 1989). 
Giovanelli \etal (1994), through a careful study of dust in 
a well-defined sample of Sc galaxies, has found that extinction
is almost non-existent two or three scale lengths away from the
center.  In M96, all but 14 of the planetaries have projected
galactocentric distances greater than three scale lengths 
(Kent 1985); in M51, virtually all of the identified
PN are well outside the main body of the galaxy.  Additionally, 
Rowan-Robinson (1992) has fit dust models
to IRAS data in many disk galaxies, and has found that the amount
of extinction averaged over the entire disk is small, with
$A_V < 0.1$~mag except near the galactic centers. 
His dust models of M101 and M51 imply mean differential 
extinctions of $E(B$$-$$V) = 0.0056$ and 0.013, respectively.
Thus, again, the extinction applicable to our PN population appears to be
small.

A third justification for neglecting the effects of internal extinction
comes from considering the positions of planetary nebulae within our own
Galaxy and other spiral galaxies.  In the Milky Way, planetary nebulae have
a higher vertical scale height than the dust (Allen 1973).  
Planetary nebulae are also seen at high $z$ distances in the edge-on 
Sb spirals NGC 891 and NGC 4565 (Ciardullo, Jacoby, \& Harris 1991; 
Jacoby, Ciardullo, \& Harris 1996).  
Therefore, it is likely that a substantial fraction
of our PN candidates lie above and below the dust layers of the
target galaxies.  Since only those planetaries below the dust will be 
extincted, a thin layer of extinction should only distort the faint 
end of the PNLF by dimming objects on the far side of the galaxy; 
the computed distance, which comes from the brightest objects in 
front of the dust, should remain unaffected.  This hypothesis can be 
tested quantitatively by first assuming that the dust and planetaries
in M101 have the same vertical scale heights as in our Milky Way galaxy,
and then modeling the PNLF expected from such a distribution.  By varying the
total amount of extinction at $\lambda 5007$ perpendicular to the disk,
we can compute a dust-altered PNLF and fit this function to our M101 data.  
The results are plotted in Figure~4.  As expected, dust has little effect on 
our PNLF distances: the derived distance modulus is always within 0.1~mag of 
the result obtained by assuming no internal extinction.  This offset 
is comparable to our internal error bars.

Finally, we can ask what the worst case scenario for dust should be.  
If all of our planetaries are Pop~I disk objects, they should  
suffer a comparable amount of extinction to that of Cepheid variables.   
In the {\sl Hubble Space Telescope\/} Cepheid observations of M101,
Kelson \etal (1996) derived a mean $E(B$$-$$V)$ from their observations
of 0.03 mag.  Using a Seaton (1979) reddening curve, the extinction in 
[O~III] would then be $A_{5007} = 0.11$ mag, similar to that estimated
above.  

In conclusion, it is likely that the effect of dust on the PNLF in face-on 
spirals is minimal, and certainly no more than $\sim  0.1$~mag.  Hence
we make no correction for internal extinction at this time.  In the future, 
this assumption can be put to a direct observational test by
observing the Balmer lines in our large sample of planetaries.

\section{6. Detection of Supernovae}

In previous PNLF studies of elliptical galaxies, searches have been made
for the remnants of historical supernovae, since these objects may contribute 
to the population of overluminous [O~III] $\lambda 5007$ sources (Jacoby, 
Ciardullo, \& Harris 1996).  Interestingly, none of these remnants have been 
detected, either because the remnants are still too dense to emit at $\lambda 
5007$, or because the local interstellar medium is so sparse that the shells 
have already expanded to an undetectable low density (Jacoby, Ciardullo, \& 
Ford 1990a).
In our sample of spiral galaxies, there have been five historical supernovae:
three in M101 (SN~1909A, 1951H, and 1970G) and two in the M51 system
(SN~1945A and 1994I).  To assess the effect of recent supernovae on the bright
end of the PNLF, we searched our data for the remnants of these
explosions.  SN 1909A was too far west of M101's nucleus to be included
on our images, but the other four supernovae are contained within our survey
fields.  We discuss each supernova below.

\subsection{6.1. SN 1970G}
SN~1970G, a type~II-L supernova, has been detected in the radio
(Gottesmann \etal 1972), and recently has been recovered in the optical
by Fesen (1993).  Fesen's spectra show little or no [O~III] emission 
related to the supernova.  The images of the on-band and off-band frames 
are displayed in Figure~5, and should be compared with the $R$-band image 
given in Fesen's paper.  As can be seen in the figure, the supernova
is detectable on the off-band image, but there is no corresponding 
emission at $\lambda 5007$.  We did not recover the supernova in
our $R$-band exposure, but since the seeing on the frame was
poor and our exposure time was much shorter than that of Fesen, the
two results are still consistent.  The SN appears to be a continuum source, 
but not a source of [O~III] $\lambda 5007$ emission.

\subsection{6.2. SN 1951H}
The type~II supernova, SN~1951H was originally discovered by Humason and
announced as a supernova by Bowen (1951).  The supernova lies just to 
the east of NGC 5462, a giant H~II region in one of the spiral arms of 
M101.  In Figure~6, we display the region surrounding NGC 5462.  From a 
visual inspection of a supernova image (Abell 1975), 
it appears that SN~1951H's reported nuclear offset of $350\arcsec$
north and $45\arcsec$ east (Barbon, Cappellaro, \& Turatto 1989) is incorrect;  
the object appears to be located approximately $8\arcsec$ east,
and $3\arcsec$ north of the reported position.  No [O~III] emission is 
seen at this location, though there does appear to be at least
two continuum sources in the area.  Further study is
not possible at this time, due to the positional uncertainty:
there are no recent optical measurements, and the object has not been 
recovered in the radio (Allen, Goss, \& van Woerden 1973).  This problem
can, however, be resolved in the future via astrometric measurements of 
the original photographic plates.

\subsection{6.3. SN 1994I}

SN~1994I, a type~Ic supernova near M51's nucleus, was discovered on 
1994 April 2, by several sets of observers. This object has been extremely 
well studied in the optical (for a summary, see Richmond \etal 1996), 
and also has a radio detection (Rupen \etal 1994).  Using images kindly 
provided to us by Michael Richmond and Schuyler Van Dyk, we searched 
for evidence of the supernova in our images.  The supernova was detected 
on all our frames, which suggests the object is still in decline from the 
initial explosion.  At the time of our survey ($\sim 1$~yr after maximum), 
there was no evidence for nebular activity.

\subsection{6.4. SN 1945A} 

SN~1945A, a type~I supernova, exploded in the satellite galaxy of
M51, NGC~5195.  The recorded position of the event was $6\arcsec$ west and 
$4\arcsec$ south of the galaxy nucleus.  No point source appears
at these coordinates, either in the continuum, or in [O~III]\null.  
Due to the lack of accurate astrometric coordinates for 
this object, we cannot draw any further conclusions. 

\subsection{6.5. Discussion}

In our sample of four historical supernovae, we detected only 
one in our [O~III] $\lambda 5007$ filter, SN~1994I, but this object is 
clearly a continuum source still in decline from the initial explosion.  
How does this agree with previous results?  Currently, there are at least 
ten historical extragalactic supernovae that have been optically recovered 3
years or more after the initial explosion (for a review, see Leibundgut 1994).
Most of these objects do have strong [O~III] emission, with SN~1970G being 
the exception.  So far, the optical emission from historical supernovae has 
always been accompanied by corresponding radio emission; this is thought to 
be due to the interaction of the supernova shock wave with circumstellar 
matter released previously in the star's lifetime (Chevalier 1982, 1984).
This naturally explains why only historical supernovae of type~II and type 
I~b/c have been recovered in the optical and radio.  Looking at our
particular cases, we see than SN~1945A was a type~I, and thus should not 
be emitting in [O~III] $\lambda 5007$.  Similarly, SN~1951H has not been 
detected in the radio, and so it, too, may not be undergoing an interaction 
with a circumstellar medium.  Finally, although SN~1970G does have a radio 
detection, the object has never been seen to have [O~III] lines, so it 
should also be undetectable in our survey.

From other studies, the rate of recovery of historical supernovae
is low.  Out of the six supernovae which have occurred in NGC~6946 in the
past 75 years, only one remnant has been detected (Fesen \& Becker 1990). 
Similarly, only one of the 6 historical supernovae of M83 has been recovered
(Long, Blair, \& Krzeminski 1989).  Thus, our non-detection of supernovae is 
consistent with other searches.

How does the presence of supernova remnants affect the PNLF as a distance
indicator?  With the exception of SN~1957D (Long, Blair, \& Krzeminski 1989), 
[O~III] $\lambda 5007$ emission is always accompanied by strong
H$\alpha$ emission.  Thus our H$\alpha$ frames should allow us to remove
supernova remnants in the same way as H~II regions.  Other ways of
discriminating luminous PN from supernova remnants include looking
for radio emission or the presence of shock diagnostic emission lines,
such as [S~II] $\lambda\lambda 6716, 6731$.  Finally, we note that [O~III]
$\lambda 5007$ has not been detected in any historical type~Ia SN event 
(Leibundgut 1994).  Thus, the PNLF in normal elliptical galaxies should 
have no contamination from supernova remnants.

\section{7. The Cepheid - PNLF Comparison}

With the addition of M101 and M96, there are now seven galaxies with 
both a PNLF and Cepheid distance determination.  In addition, there are 
three galaxy groups where an indirect comparison can be made: the NGC~1023
Group (Silbermann \etal 1996; Ciardullo, Jacoby, \& Harris 1991), the Virgo 
Cluster (Ferrarese \etal 1996; Saha \etal 1996; Jacoby, Ciardullo, \& Ford
1990a), and the Fornax Cluster (Freedman 1996; McMillan, Ciardullo, \& Jacoby 
1993).  In Figure~7 and Table~8, we compare the 
PNLF and Cepheid distances to these objects.  The error bars given in the
table are those quoted by the authors, with the systematic uncertainties 
common to both methods removed.  For the PNLF technique, these systematic 
uncertainties include the 0.1~mag error associated with the distance to M31, 
and the 0.05~mag uncertainty in the definition of the PNLF; for the Cepheid 
measurements, the $\sim 0.1$~mag uncertainty associated with the distance and 
reddening to the Magellanic Clouds has been excluded.  To place the PNLF 
on the same scale as the Cepheids, we have adopted an M31 distance of 770~kpc 
(Freedman \& Madore 1990), and a foreground reddening of $E(B$$-$$V) = 0.08$ 
(Burstein \& Heiles 1984). In the case of NGC 5253 only, a correction to
its distance might be needed, if the galaxy is extremely metal poor
(Ciardullo \& Jacoby 1992).  There is evidence that the metallicity
of NGC 5253 (Melnick \etal 1992; Pagel \etal 1992; Walsh \& Roy 1989) 
warrants a correction as large as $-0.28$ mag, but since others (e.g., Campbell
1992; Storchi-Bergmann \etal 1994) report near-LMC metallicities, 
it is unclear whether this correction should be used.  We have chosen
not to apply a distance correction to NGC~5253; were we to do so,
the discrepancy between the PNLF and Cepheid distances to the galaxy
would change from $+0.4 \, \sigma$ to $-0.5 \, \sigma$. 

As can be seen in the figure, there is extremely good agreement over two 
orders of magnitude in the derived distances: the only galaxy whose Cepheid 
distance varies by more than $1\sigma$ from the PNLF distance is M96.  In the 
bottom of Figure~7, we plot the fractional residuals between the 
Cepheid and PNLF distances.  Note that, with the exception of M96, the 
scatter about the zero line is completely consistent with the error bars.  
Moreover, if we were to adopt the Leo~I Group distance of Graham \etal (1996),
rather than that of Tanvir \etal (1995), M96, too, would 
agree within the errors.  

The data presented in Table~8 affords us an opportunity to check the
zero point calibration of the PNLF by computing the mean value of 
$\Delta = (m-M)_{\rm PNLF} - (m-M)_{\rm Cepheid}$ for our sample of galaxies.
From the six galaxies (other than M31) with both Cepheid and PNLF distance
measurements, we find $\langle \Delta \rangle$ of $-0.01$ (weighted)
and $\langle \Delta \rangle = +0.01$ (unweighted).  If the Graham \etal 
Cepheid distance to M95 is substituted for the Tanvir \etal M96 distance, 
this offset changes to $\langle \Delta \rangle = +0.03$ (weighted) and 
$\langle \Delta \rangle = +0.06$ (unweighted).   Finally, if the results 
from the NGC~1023 Group, the Fornax Cluster, and the Virgo Cluster are 
folded in, with an assumed intracluster scatter  of $\pm 1$ Mpc, 
then $\langle \Delta \rangle = +0.02$ (weighted) and 
$\langle \Delta \rangle = +0.03$ (unweighted).  This excellent agreement is 
yet another confirmation of the robustness of the PNLF method, and 
demonstrates that our calibration of $M^*$ in M31 is certainly within the 
quoted errors.

The agreement between the PNLF and Cepheid distances also 
places a possible constraint on theories for the dependence of the
planetary nebula luminosity function with population age.  Models by Jacoby 
(1989), M\'endez \etal (1993), and Han, Podsiadlowski, \& Eggleton (1994) 
all suggest that a population of PN derived from young, 
massive stars, should have a value of $M^*$ brighter than that seen in an old 
stellar population.  Specifically, M\'endez \etal (1993) predict a difference 
of $\sim 0.6$~mag between the PNLF cutoff in a population with a constant star 
formation rate, and that in an old elliptical galaxy.  

This effect is not seen in our data, except possibly in M96.  Indeed, in 
the case of M101, if there is any effect at all, it goes in the wrong 
direction: our distance to M101 is slightly larger than that derived from 
the Cepheids, and this implies a slightly fainter value for $M^*$.  
A possible explanation for the apparent lack of an age dependence
may lie in our selection criteria for planetary nebulae.  In order to avoid 
contamination by H~II regions, our PN survey preferentially identified 
objects away from star forming regions.  Consequently, we may be 
discriminating against the highest mass PN\null.  In addition, the (small) 
effects of internal extinction within the surveyed spirals may also work to
mask a population effect.  Nevertheless, the results are consistent with
the PNLF analysis of Jacoby, Walker, \& Ciardullo (1990b) 
for the Large Magellanic Cloud.  No age effect was seen in that 
galaxy, either.  If this lack of an age effect is real, and not due to a 
selection effect, it implies either that a) the PN progenitors 
in late-type spirals have similar ages to 
those in ellipticals and spiral bulges, b) the initial-to-final mass 
relationship for post-asymptotic branch stars is nearly independent 
of progenitor mass, or c) high mass progenitors conspire to be 
relatively faint in [O~III] $\lambda 5007$, perhaps due to their 
short evolutionary time scales (\eg Sch\"onberner 1981; 1983), high 
nitrogen-to-oxygen abundance ratio (Kaler \& Jacoby 1991), 
or differences in geometry (M\'endez 1996).  However, more tests with
spiral galaxies will be needed to test this hypothesis.  

\section{8. Conclusion}

We have searched for planetary nebulae in the spiral galaxies M101, M51
and M96, and obtained distance moduli of $(m$$-$$M)_0 = 29.42 \pm 0.15$, 
$(m$$-$$M)_0 = 29.62 \pm 0.15$, and $(m$$-$$M)_0 = 29.91 \pm 0.15$
respectively, using the planetary nebula luminosity function. 
The M101 distance agrees extremely well with the {\sl HST\/} Distance Scale
Key Project Cepheid measurement, and the M51 distance is in excellent
agreement with the distance determined by surface brightness fluctuations.  
The M96 distance is consistent with other PNLF measurements in the Leo~I 
group, and the Leo~I distance found by the {\sl HST\/} Distance Scale Key
Project, but disagrees with the Cepheid distance found by Tanvir {\it
et al.}  The success of our observations implies that PNLF measurements are 
an effective ground-based alternative to {\sl Hubble Space Telescope\/}
measurements of Cepheid light curves. 

We thank M. Bershady and R. Wade for comments on an earlier version
of this paper, and E. Baron for comments on the SEAM distance
to M51.  We especially thank M. Richmond and S. van Dyk for
the use of their SN~1994I images in our supernova search.  We also 
thank the referee, L. Stanghellini, for several suggestions that 
improved this paper.  This work was supported in part by 
NASA grant NAGW-3159.

\vfill\eject

$$\vbox{
\catcode`?=\active
\def?{\kern\digitwidth}
\halign {#\hfil&#\hfil&\hfil#\hfil&\hfil#\hfil&\hfil#\hfil\cr
\multispan{5} \hfil TABLE 1 \hfil\cr
\noalign {\vskip5pt}
\multispan{5} \hfil Record of Observations \hfil\cr
\noalign {
 \vskip3pt 
 \hrule height1pt 
 \vskip2pt 
 \hrule height1pt 
 \vskip6pt
}
            &      &        & Exposure &Number of \cr
Object Name & Date & Filter & (sec)    & Exposures\cr
\noalign {\vskip3pt\hrule height1pt \vskip6pt}
M 101 & 1995 April 5  & 5027/30   & 3600   & 2 \cr
M 101 & 1995 April 5  & 5312/267  & ?540   & 2 \cr
M 101 & 1995 April 5  & 6586/72   & ?900   & 1 \cr
M 96  & 1995 April 6  & 5027/30   & 3600   & 3 \cr
M 96  & 1995 April 6  & 5312/267  & ?540   & 3 \cr
M 96  & 1995 April 7  & 6586/72   & ?900   & 3 \cr
M 96  & 1995 April 7  & 5312/267  & ?540   & 2 \cr
M 51  & 1995 April 7  & 5027/30   & 3600   & 2 \cr
M 51  & 1995 April 7  & 5312/267  & ?540   & 2 \cr
M 101 & 1995 April 8  & 6563/75   & ?900   & 4 \cr
M 101 & 1995 April 8  & $R$ band  & ??60   & 4 \cr
M 101 & 1995 April 11 & 6563/75   & ?900   & 2 \cr
M 101 & 1995 April 11 & $R$ band  & ??60   & 2 \cr
M 51  & 1995 April 11 & 6563/75   & ?900   & 6 \cr
M 51  & 1995 April 11 & R band    & ??60   & 6 \cr
\noalign {\vskip8pt\hrule height1pt}
}
}$$
\vfill\eject

\epsfbox[50 288 570 756]{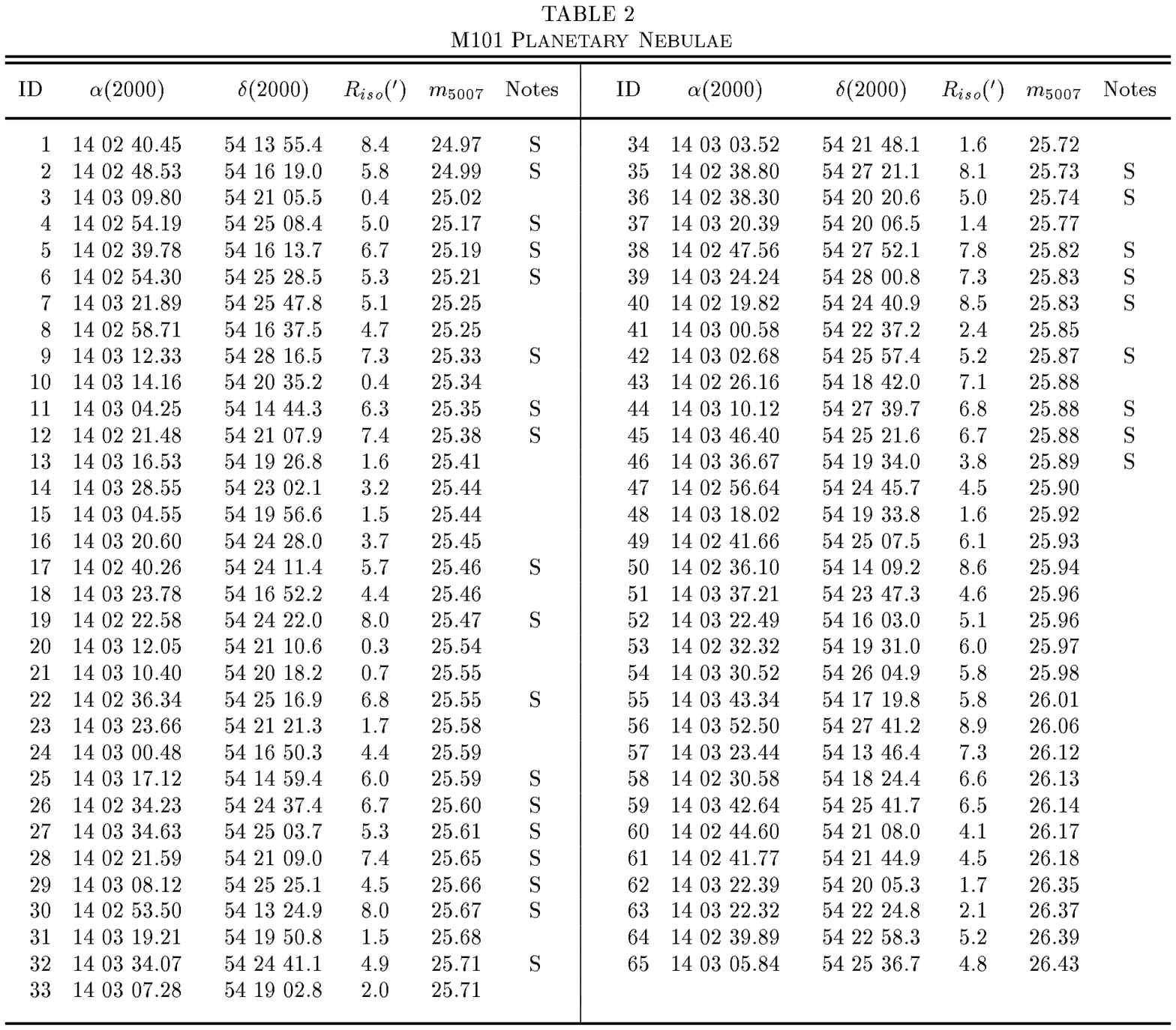}

\vfill\eject

\epsfbox[50 300 570 756]{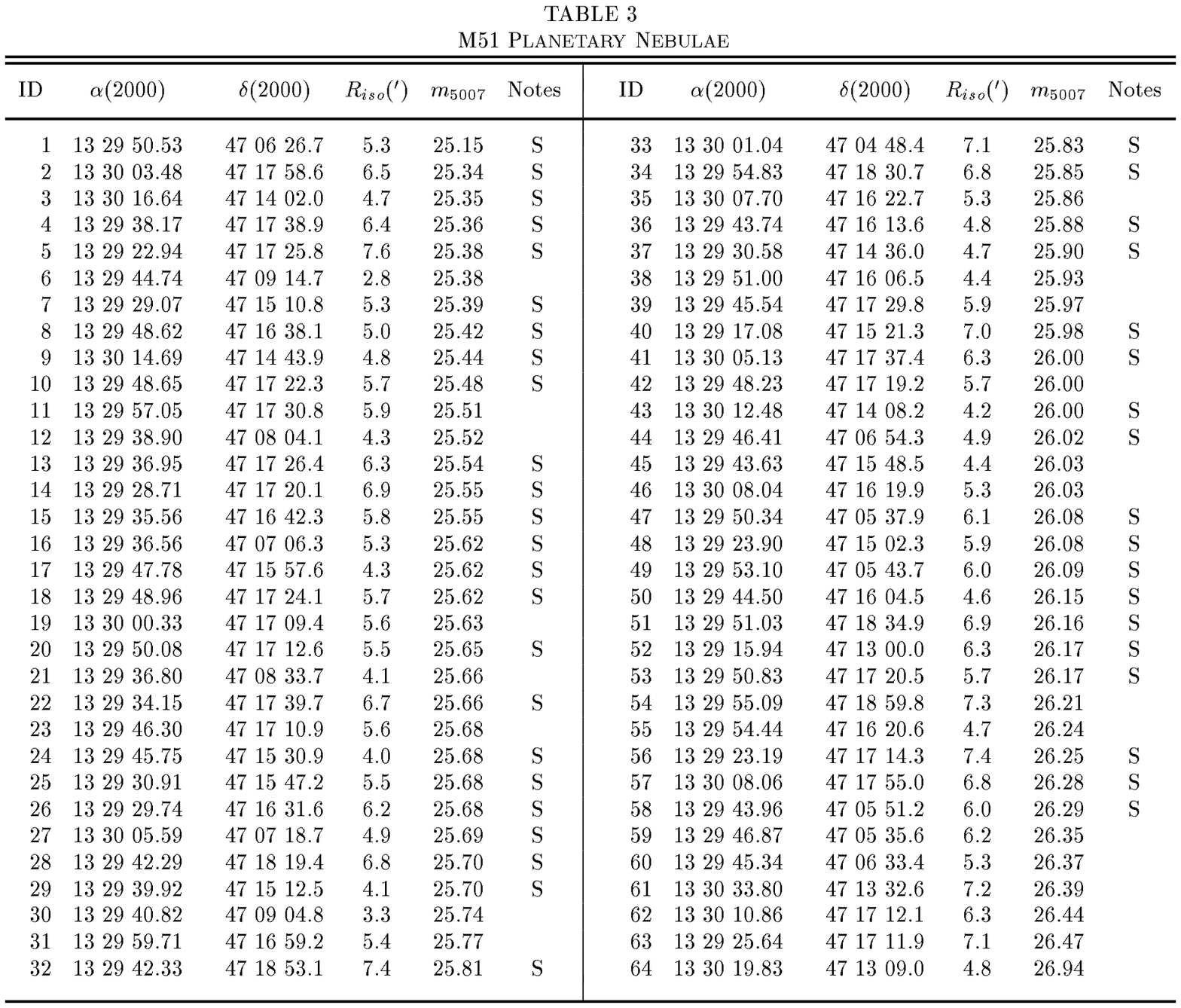}

\vfill\eject

\epsfbox[50 240 570 756]{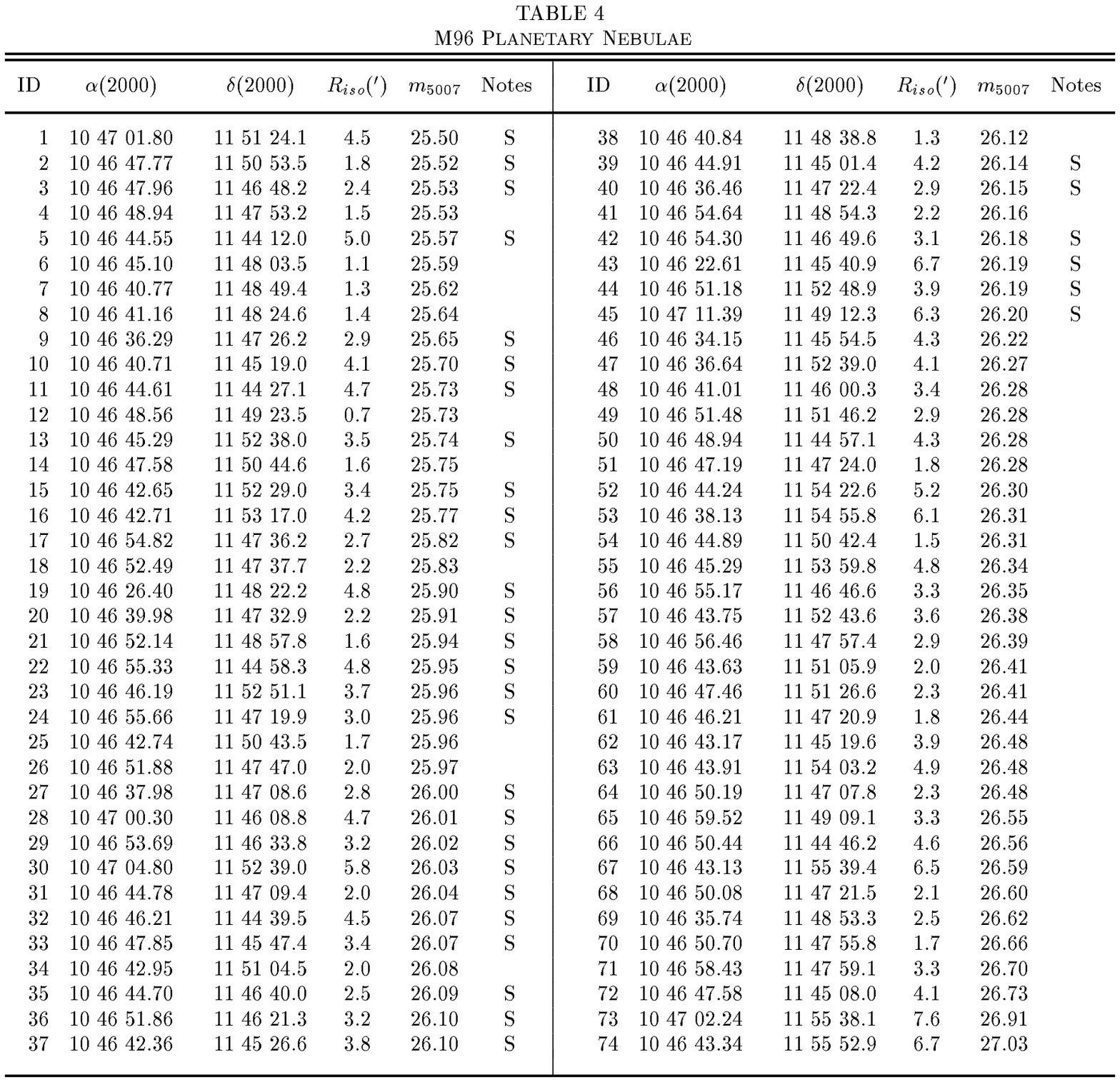}

\vfill\eject

$$\vbox{
\catcode`?=\active
\def?{\kern\digitwidth}
\halign {#\hfil&#\hfil&#\hfil&\hfil#\hfil&\hfil#\hfil\cr
\multispan{5} \hfil TABLE 5 \hfil\cr
\noalign {\vskip5pt}
\multispan{5} \hfil Distances to M101 \hfil\cr
\noalign {
 \vskip3pt 
 \hrule height1pt 
 \vskip2pt 
 \hrule height1pt 
 \vskip6pt
}
Year &Author &Method and Data &$(m$$-$$M)_0$ &$\sigma$ \cr
\noalign {\vskip3pt\hrule height1pt \vskip6pt}
1974a &Sandage \& Tammann &Multiple Methods &29.3? &0.3 \cr
1975  &de Vaucouleurs &Group Membership &28.71\cr
1976  &Sandage \& Tammann &Multiple Methods & 29.08 &0.3 \cr
1976  &Jaakkola \& Le Denmat &Revision of S\&T (1974b, 1976) &28.56 \cr
1976  &Bottinelli \& Gouguenheim &Revision of S\&T (1974b, 1976)  &28.75 \cr
1977  &Melnick &Velocity Dispersion in H~II Regions &28.7? & 0.6\cr
1978  &de Vaucouleurs &Group Membership & 28.5? &0.3 \cr
1979a &de Vaucouleurs &Revision of Melnick (1977) &28.08 &0.4 \cr
1979b &de Vaucouleurs &Luminosity index &27.57 &0.3\cr
1979b &de Vaucouleurs &Type II SNe &27.60 \cr
1980  &Wray \& de Vaucouleurs &Brightest Superassociation &27.76 &0.6\cr
1980  &Capaccioli \& Fasano &Revision of S\&T (1974b, 1976) &28.8? &0.3 \cr
1982  &Lawrie \& Kwitter  &Largest H~II ring &28.18 &0.28 \cr
1983  &Sandage & Brightest blue irreg.~variables &29.2? \cr
1983  &Sandage & Brightest red supergiants &29.2? \cr
1983  &Humphreys \& Strom &Brightest M Supergiants &28.6? &0.3\cr
1984  &Bottinelli \etal &B-band T-F relation &28.7? \cr
1985  &Bottinelli \etal &B-band T-F relation &28.4? \cr
1986  &Humphreys \etal &Brightest M Supergiants &28.4? \cr
1986  &Cook \etal &R-Band Cepheids &29.2? &0.1 \cr
1992  &Schmidt \etal &EPM - SN 1970G &29.34 &$^{+0.28}_{-0.49}$ \cr
1994  &Pierce & $B$$R$$I$-Band T-F relation &29.2? &0.5\cr
1995  &Alves \& Cook &Cepheids \& Mira stars &29.08 &0.13\cr
1996  &Kelson \etal & HST Cepheids &29.34 &0.17\cr
1996  &Feldmeier \etal &PNLF &29.42 &0.15\cr
\noalign {\vskip8pt\hrule height1pt}
}
}$$
\vfill\eject

$$\vbox{
\catcode`?=\active
\def?{\kern\digitwidth}
\halign {#\hfil&#\hfil&#\hfil&\hfil#\hfil&\hfil#\hfil\cr
\multispan{5} \hfil TABLE 6 \hfil\cr
\noalign {\vskip5pt}
\multispan{5} \hfil Distances to M51 \hfil\cr
\noalign {
 \vskip3pt
 \hrule height1pt 
 \vskip2pt 
 \hrule height1pt 
 \vskip6pt
}
Year &Author &Method and Data &$(m$$-$$M)_0$ &$\sigma$ \cr
\noalign {\vskip3pt\hrule height1pt \vskip6pt}
1974b &Sandage \& Tammann &H~II Regions &29.91 \cr
1990  &Georgiev \etal  &Stellar Association Sizes &29.2? &0.2\cr
1994  &Iwamoto \etal  &SN 1994I - C+O stars &29.2? &0.2\cr
1996  &Baron \etal &SN 1994I - SEAM &28.9? &0.6\cr
1996  &Tonry     &SBF  &29.59 &0.15\cr
1996  &This work &PNLF &29.62 &0.15\cr
\noalign {\vskip8pt\hrule height1pt}
}
}$$
\vfill\eject

$$\vbox{
\catcode`?=\active
\def?{\kern\digitwidth}
\halign {#\hfil&#\hfil&#\hfil&\hfil#\hfil&\hfil#\hfil\cr
\multispan{5} \hfil TABLE 7 \hfil\cr
\noalign {\vskip5pt}
\multispan{5} \hfil Distances to M96 (Including Leo Group) \hfil\cr
\noalign {
 \vskip3pt
 \hrule height1pt
 \vskip2pt
 \hrule height1pt
 \vskip6pt
}
Year &Author &Method and Data &$(m$$-$$M)_0$ &$\sigma$ \cr
\noalign {\vskip3pt\hrule height1pt \vskip6pt}
1977  &Visvanathan \& Sandage &Color-Magnitude relation &31.18 \cr
1979  &Visvanathan &Color-Magnitude relation &31.09 &0.33\cr
1982  &de Vaucouleurs \& Olson &Faber-Jackson relation &29.98 &0.26\cr
1983  &Aaronson \& Mould &Tully Fisher (on Leo Triplet) &29.84 &0.25\cr
1985  &Pritchet \& van den Bergh &Globular clusters LF &29.16 &0.3\cr
1988  &Tonry \& Schneider &Luminosity Fluctuations &29.96 \cr
1989b &Ciardullo \etal &PNLF (group) &30.02 &0.10\cr
1990  &Harris      &Globular Clusters LF &30.19 &0.43\cr
1991  &Tonry &Luminosity fluctuations (group) &29.84 \cr
1995  &Tanvir \etal &HST Cepheids &30.32 &0.16\cr
1996  &Sakai \etal &Tip of RGB &30.20 &0.14\cr
1996  &Graham \etal &HST Cepheids &30.01 &0.19\cr
1996  &This work &PNLF &29.91 &0.15\cr
\noalign {\vskip8pt\hrule height1pt}
}
}$$
\vfill\eject

$$\vbox{
\catcode`?=\active
\def?{\kern\digitwidth}
\halign {#\hfil&#\hfil&\hfil#\hfil&#\hfil&\hfil#\hfil\cr
\multispan{5} \hfil TABLE 8 \hfil\cr
\noalign {\vskip5pt}
\multispan{5} \hfil Galaxies with PNLF and Cepheid Distances \hfil\cr
\noalign {
 \vskip3pt
 \hrule height1pt
 \vskip2pt
 \hrule height1pt
 \vskip6pt
}
Galaxy &Fitted Objects &Distance &Reference &$\Delta$\cr
\noalign {\vskip3pt\hrule height1pt \vskip6pt}
M31 &104 PNe &\dots &Ciardullo \etal 1989a \cr
&38 Cepheids &$24.42 \pm 0.10$ &Freedman \& Madore 1990 &\dots\cr
\cr
LMC &58 PNe & $18.49^{+0.12}_{-0.14}$ &Jacoby \etal 1990b \cr
&28 Cepheid & $18.47 \pm 0.10$ &Feast \& Walker 1987 &$+0.02^{+0.20}_{-0.21}$\cr
\cr
NGC 300 &25 PNe &$26.84^{+0.21}_{-0.29}$ &Soffner \etal 1996\cr
&10 Cepheids & $26.66 \pm 0.10$ &Freedman \etal 1992 &$+0.27^{+0.23}_{-0.31}$\cr
\cr
M81 &89 PNe &$27.78^{+0.08}_{-0.09}$ &Jacoby \etal 1989 \cr 
&31 Cepheids &$27.80 \pm 0.17$ &Freedman \etal 1994  &$-0.02^{+0.19}_{-0.19}$\cr
\cr
NGC 5253$^a$ &10 PNe &$28.14^{+0.14}_{-0.46}$ &Phillips \etal 1992\cr	
&12 Cepheids &$27.94 \pm 0.10$  &Saha \etal 1995  &$+0.20^{+0.17}_{-0.47}$\cr
\cr
M101 &27 PNe   &$29.42^{+0.09}_{-0.12}$ &Feldmeier \etal 1996 \cr
&29 Cepheids   &$29.34 \pm 0.08$ &Kelson \etal 1996 &$+0.08^{+0.12}_{-0.14}$\cr
\cr
M96 &33 PNe   &$29.91^{+0.08}_{-0.10}$ &This paper \cr
&7 Cepheids  &$30.32 \pm 0.13$ &Tanvir \etal 1995 &$-0.41^{+0.15}_{-0.16}$\cr
&45 Cepheids &$30.01 \pm 0.16$ &Graham \etal 1996 &$-0.10^{+0.18}_{-0.19}$\cr
\noalign {\vskip8pt\hrule height1pt}
}
}$$
$^a$The quoted distances to NGC~5253 assume a foreground
galactic absorption of $A_B = 0.19$ from Burstein \& Heiles (1984).
\section{References}
\beginrefs

Aaronson, M., \& Mould, J. 1983, \apj, 265, 1

Abell, G.O. 1975, {\it Exploration of the
Universe,} (New York: Holt, Rinehart, \& Wilson), p.~614

Allen, C.W. 1973, {\it Astrophysical 
Quantities, 3rd edition,} (London: Athlone Press)

Allen, R.J., Goss, W.M., \& van Woerden, H. 1973, \aap, 29, 447

Alves, D.R., \& Cook, K.H. 1995, \aj, 110, 192

Barbon, R., Cappellaro, E., \& Turatto, M. 1989, \aaps, 81, 421

Baron, E. 1996, private communication.

Baron E., Hauschildt, P.H., Branch, D., 
Kirshner, R.P., \& Filippenko, A.V. 1996, \mnras, 279, 799

Bottinelli, L., \& Gouguenheim, L. 1976, \aap, 51, 275

Bottinelli, L., Gouguenheim, L., Paturel, G., \& de Vaucouleurs, G. 
1984, \aaps, 56, 381

Bottinelli, L., Gouguenheim, L., Paturel, G., \& de Vaucouleurs, G.
1985, \aaps, 59, 43

Bottinelli, L., Gouguenheim, L., Paturel, G., \&
Teerikorpi, P. 1991, \aap, 252, 550

Bowen, I.S. 1951, {\it Carnegie Yrb., \bf 50,} 17

Burkhead, M.S. 1978, \apjs, 38, 147

Burstein, D., \& Heiles C. 1984, \apjs, 54, 33

Buta, R., Corwin, H.G., de Vaucouleurs, G., de Vaucouleurs, A., \&
Longo, G. 1995, \aj, 109, 517

Campbell, A. 1992, \apj, 401, 157
 
Capaccioli, M., \& Fasano, G. 1980, \aap, 83, 354

Chevalier, R.A. 1982, \apj, 251, 259

Chevalier, R.A. 1984, {\it Ann.~N.Y.~Acad.~Sci., \bf 422,} 215

Ciardullo, R. 1995, {\it IAU Highlights of Astronomy, 10,}
ed.~I. Appenzeller (Dordrecht: Kluwer), p.~507

Ciardullo, R., Ford, H.C., Neill, J.D., Jacoby, G.H., \& Shafter, A.W. 
1987, \apj, 318, 520

Ciardullo, R., Jacoby, G.H., \& Feldmeier J.J. 1997, {\it Ap.~J.,}
in preparation

Ciardullo, R., Jacoby, G.H., \& Ford, H.C. 1989b, \apj, 344, 715

Ciardullo, R., Jacoby, G.H., Ford, H.C., \& Neill, J.D. 1989a, \apj, 339, 53

Ciardullo, R., Jacoby, G.H., \& Harris, W.E. 1991, \apj, 383, 487

Ciardullo, R., Jacoby, G.H., \& Harris, W.E. 1996, \apj, 462, 1

Ciardullo, R., Jacoby, G.H., \& Tonry, J.L. 1993, \apj, 419, 479

Cohen, J.G. 1993, \baas, 25, 818

Cook, K.H., Aaronson, M., \& Illingworth, G. 1986, \apj, 345, 245

de Vaucouleurs, G. 1958, \apj, 128, 465

de Vaucouleurs, G. 1975, {\it Stars and Stellar Systems IX,
Galaxies and the Universe,} ed.~A. Sandage, M. Sandage, \& J. Kristian
(Chicago: University of Chicago Press), p.~557

de Vaucouleurs, G. 1978, \apj, 224, 710

de Vaucouleurs, G. 1979a, \aap, 79, 274

de Vaucouleurs, G. 1979b, \apj, 227, 729

de Vaucouleurs, G. 1993, \apj, 415, 10

de Vaucouleurs, G., de Vaucouleurs,
A., Corwin, H.G., Buta, R.J., Fouqu\'e, P., \& Paturel, G. 1991, {\it Third
Reference Catalogue of Bright Galaxies,} (New York: Springer-Verlag)

de Vaucouleurs, G., \& Olson, D.W. 1982, \apj, 256, 346

Disney, M.J., Davies, J.I., \& Phillips, S. 1989, \mnras, 239, 939

Feast, M.W., \& Walker, A.R. 1987, \araa, 25, 345

Feldmeier, J.J., Ciardullo, R., \& Jacoby, G.H. 1996, \apjl, 461, L25

Ferrarese, L., Freedman, W.L., Hill, R.J., Saha, A., 
Madore, B.F., Kennicutt, R.C., Stetson, P.B., Ford, H.C., Graham, J.A., 
Hoessel, J.G., Han, M., Huchra, J., Hughes, S.M., Illingworth, G.D.,
Kelson, D., Mould, J.R., Phelps, R., Silbermann, N.A., Sakai, S., 
Turner, A., Harding, P., \& Bresolin, F. 1996, \apj, 464, 568

Fesen, R.A. 1993, \apjl, 413, L109

Fesen, R.A., \&  Becker, R.H. 1990, \apj, 351, 437

Freedman, W.L. 1996, in {\it Proceedings of the Space 
Telescope Science Institute Symposium on the Extragalactic Distance Scale,} 
in press

Freedman, W.L., Hughes, S.M., Madore, B.F., Mould, J.R.,
Lee, M.G., Stetson, P., Kennicutt, R.C., Turner, A., Ferrarese, L.,
Ford, H., Graham, J.A., Hill, R., Hoessel, J.G., Huchra, J., \&
Illingworth, G.D. 1994, \apj, 427, 628

Freedman, W.L., \& Madore, B.F. 1990, \apj, 365, 186

Freedman, W.L., Madore, B.F., Hawley, S.L., Horowitz, I.K.,
Mould, J., Navarrete, M., \& Sallmen, S. 1992, \apj, 396, 80

Geller, M.J., \& Huchra, J.P. 1983, \apjs, 52, 61

Georgiev, Ts.B., Getov, R.G., Zamanova, V.I., \&
Ivanov, G.R. 1990, {\it Pis.~Astron.~Zh., \bf 16,} 979  

Giovanelli, R., Haynes, M.P., Salzer, J.J., Wegner, G.,
Da Costa, L.N., \& Freudling, W. 1994, \aj, 107, 2036

Gottesmann, S.T., Broderick, J.J., Brown, R.L., Balick, B.,
   \& Palmer, P. 1972, \apj, 174, 383

Graham, J.A., Phelps, R.L., Freedman, W.L., Saha, A., Stetson, P.B., 
   Madore, B.F., Silbermann, N.A., Sakai, S., Kennicutt, R.C., Harding, P., 
   Turner, A., Mould, J.R., Ferrarese, L., Ford, H.C., Hoessel, J.G., Han, 
   M., Huchra, J.P., Hughes, S.M., Illingworth, G.D., \& Kelson, D.D. 1996,
   submitted to ApJ 
 
Han, Z., Podsiadlowski, P., \& Eggleton, P.P. 1994, \mnras, 270, 121

Harris, W.E. 1990, \pasp, 102, 966

Holmberg, E. 1964, {\it Ark.~f.~Astr., \bf 3,} 387

Howard, S., \& Byrd, G.G. 1990, \aj, 99, 1798

Huchra, J.P., \& Geller, M.J. 1982, \apj, 257, 423
 
Hui, X., Ford, H.C., Ciardullo, R., \& Jacoby, G.H. 1993, \apj, 414, 463

Humason, M.L.,  Mayall, N.U., \& Sandage, A. 1956, \aj, 61, 97
 
Humphreys, R.M., \& Aaronson, M. 1987, \apj, 318, L69

Humphreys, R.M.,  Aaronson, M., Lebofsky, M., McAlary, C.W.,
Strom, S.E., \& Capps, R.W. 1986, \aj, 91, 808

Humphreys, R.M., \& Strom, S.E. 1983, \apj, 264, 458

Iwamoto, K., Nomoto, K., Hoflich, P., Yamaoka, H., Kumagai, S.,
   \& Shigeyama, T. 1994, \apjl, 437, 115

Jaakkola, T., \& Le Denmat, G. 1976, \mnras, 176, 307

Jacoby, G.H. 1989, \apj, 339, 39

Jacoby, G.H., Branch, D.,
  Ciardullo, R., Davies, R.L., Harris, W.E., Pierce, M.J., Pritchet, 
  C.J., Tonry, J.L., \& Welch, D.L. 1992, \pasp, 104, 599

Jacoby, G.H., Ciardullo, R., \& Ford, H.C. 1990a, \apj, 356, 332

Jacoby, G.H., Ciardullo, R., Ford, H.C., \& Booth, J. 1989, \apj, 344, 70

Jacoby, G.H., Ciardullo, R., \& Harris, W.E. 1996, \apj, 462, 1

Jacoby, G.H., Quigley, R.J., \& Africano, J.L. 1987, \pasp, 99, 672

Jacoby, G.H., Walker, A.R., \& Ciardullo, R. 1990b, \apj, 365, 471

Kaler, J.B., \& Jacoby, G.H. 1991, \apj, 382, 134

Kelson, D.D., Illingworth, G.D., Freedman, W.L., Hill, R.,
   Graham, J.A., Stetson, P.B., Saha, A., Madore, B.F., Kennicutt, R.C.,
   Mould, J.R., Hughes, S.M., Ferrarese, L., Phelps, R., Turner, A.,
   Cook, K.H., Ford, H.C., Hoessel, J., \& Huchra, J. 1996, \apj, 463, 26

Kent, S.M. 1985, \apjs, 59, 115

Lawrie, D.G., \& Kwitter, K.B. 1982, \apj, 255, L29

Leibundgut, B. 1994, in {\it Circumstellar
   Media in Late Stages of Stellar Evolution,} ed.~R. Clegg, P. Meikle, \&
   I. Stevens (Cambridge: Cambridge Univ.~Press), p.~100

Long, K.S., Blair, W.P., \&  Krzeminski, W. 1989, \apjl, 340, L25

Madore, B.F., \& Freedman, W.L. 1991, \pasp, 103, 933
 
McClure, R.D., \& Racine, R. 1969, \aj, 74, 1000

McMillan, R., Ciardullo, R., \& Jacoby, G.H. 1993, \apj, 416, 62

Melnick, J. 1977, \apj, 213, 15

Melnick, J., Heydari-Malayeri, M. \& Leisy, P. 1992, \aap, 253, 16

M\'endez, R.H. 1996, in {\it I.A.U. Symposium 180, Planetary
Nebulae,} ed.~H. Habing \& H. Lamers (Dordrecht: Kluwer), in press 
 
M\'endez, R.H., Kudritzki, R.P., 
    Ciardullo, R., \& Jacoby, G.H. 1993, \aap, 275, 534

Pagel, B.E.J., Simonson, E.A., Terlevich, R.J., \& 
Edmunds M.G. 1992, \mnras, 255, 325

Phillips, M.M., Jacoby, G.H.,
    Walker, A.R., Tonry, J.L., \& Ciardullo, R. 1992,  \baas, 24, 79

Pierce, M.J. 1994, \apj, 430, 53

Pritchet, C., \& van den Bergh, S. 1985, \jrasc, 79, 240

Renzini, A., \& Buzzoni, A. 1986, in {\it Spectral Evolution 
   of Galaxies,} ed.~C. Chiosi, \& A. Renzini (Dordrecht: Reidel), p.~195

Richmond, M.W., Van Dyk, S.D.,
    Ho, W., Peng, C., Paik, Y., Teffers, R.R., \& Filippenko, A.V. 1996,
    \aj, 111, 327

Rowan-Robinson, M. 1992, \mnras, 258, 787

Rozanski, R., \& Rowan-Robinson, M. 1994, \mnras, 271, 530
 
Rupen, M.P., Sramek, R.A., Van Dyk, S.D., Weiler, K.W., \& Panagia, N. 1994,
{\it IAU Circ.} No. 5963 

Saha, A., Sandage, A., Labhardt, L., Schwengeler, H., 
   Tammann, G.A., Panagia, N., \& Macchetto, F.D. 1995, \apj, 438, 8

Saha, A., Sandage, A., Labhardt, L., Tammann, G.A., 
   Macchetto, F.D., \& Panagia, N. 1996, \apjpress

Sakai, S., Madore, B.F., Freedman, W.L., Lauer, T.R., 
Ajhar, E.A., \& Baum, W.A. 1996, \apjpress

Sandage, A. 1983, \aj, 88, 1569
	
Sandage, A., \& Tammann, G.A. 1974a, \apj, 194, 223

Sandage, A., \& Tammann, G.A. 1974b, \apj, 194, 559

Sandage, A., \& Tammann, G.A. 1976, \apj, 210, 7

Sandage, A., \& Tammann, G.A. 1987,
{\it A Revised Shapley-Ames Catalog of Bright Galaxies,} 
(Washington: Carnegie Institution of Washington)

Schmidt, B.P., Kirshner, R.P., \& Eastman, R.G. 1992, \apj, 395, 366

Schneider, S.E. 1989, \apj, 343, 94

Schneider, S.E., Skrutskie, M.F., Hacking, P.B., 
Young, J.S., \& Dickman, R.L. 1989, \aj, 97, 666

Sch\"onberner, D. 1981, \aap, 103, 119

Sch\"onberner, D. 1983, \apj, 272, 708

Shaver, P.A., McGee, R.X., Newton, L.M., Danks, A.C.,
\& Pottasch S.R. 1983, \mnras, 204, 53

Seaton, M.J. 1979, \mnras, 187, 73p

Silbermann, N.A., Harding, P., Madore, B.F., Kennicutt, R.C.,
   Saha, A., Stetson, P.B., Freedman, W.L., Mould, J.R., Graham, J.A.,
   Hill, R.J., Turner, A., Bresolin, F., Ferrarese, L., Ford, H., 
   Hoessel, J.G., Han, M., Hughes, S.M.G., Illingworth, G.D., Phelps, R.,
   \& Sakai, S. 1996, \apjpress

Soffner, T., M\'endez, R.H., Jacoby, G.H., Ciardullo, R., Roth, M.M.,
   \& Kudritzki, R.P. 1996, \aap, 306, 9

Stanghellini, L. 1995, \apj, 452, 515

Stetson, P.B. 1987, \pasp, 99, 191

Stone, R.P.S. 1977, \apj, 218, 767

Storchi-Bergmann, T., Calzetti, D., \& Kinney, A.L. 1994, 
\apj, 429, 572

Tanvir, N.R., Shanks, T., Ferguson, H.C., \& Robinson, D.R.T. 1995, 
   \nat, 377, 27

Tonry, J.L. 1991, \apjl, 373, L1

Tonry, J.L. 1996, private communication

Tonry, J.L., Ajhar, E.A., \& Luppino, G.A. 1990, \aj, 100, 1416

Tonry, J.L., \& Schneider, D.P. 1988, \aj, 96, 807

Toomre, A., \& Toomre, J. 1972, \apj, 178, 623

Tully, R.B. 1988, {\it Nearby Galaxies Catalog,} 
   (Cambridge: Cambridge University Press)

Turner, E.L., \& Gott, J.R. 1976, \apjs, 32, 409

Valentijn, E.A. 1990, \nat, 346, 153

Vennik, J. 1984, Tartu Astr.~Obs., Teated, 73, 3

Visvanathan, N. 1979, {\it Ast.~Soc.~Aus.~Proc., \bf 3,} 309
   
Visvanathan, N., \& Sandage, A. 1977, \apj, 216, 214

Walsh, J.R., \& Roy, J.R. 1989, \mnras, 239, 297

Wray, J.D., \& de Vaucouleurs, G. 1980, \aj, 85, 1

Welch, D.L., McAlary, C.W., McLaren, R.A., \& Madore, B.F. 1986, \apj, 305, 583
 
\endrefs
\vfill\eject
\section{Figure Captions}  
\begincaptions

Figure 1: The planetary nebula luminosity functions for M101, M51, and M96 
binned into 0.2~mag, 0.2~mag, and 0.15~mag intervals respectively. 
The solid lines represent the empirical PNLF of equation (1) 
convolved with the mean photometric error vs.~magnitude relation 
and translated to the most likely distance modulus for each galaxy.  The
solid circles represent objects in our statistical PN samples; the open
circles indicate objects fainter than the completeness limit that were not 
included in the maximum likelihood solution.  

Figure 2:  An [O~III] $\lambda 5007$ image of M51, with the positions of our PN 
candidates marked by plus signs.  North is up, and east is to the left; the 
image is $16\arcmin$ on a side.  Note the large number of planetary 
nebulae to the west of NGC~5195. This region corresponds to the increase in 
surface brightness seen in the deep images of Burkhead (1978). 

Figure 3:  An [O~III] $\lambda 5007$ image of M96, with the positions of our PN 
candidates marked by plus signs.  North is up, and east is to the left; the 
image is $16\arcmin$ on a side.  Again, the planetaries are found most easily 
in the outskirts of the galaxy.

Figure 4:  The effect of internal extinction
on our derived distance to M101, assuming the scale height of
dust and PN in the galaxy is the same as that in Milky Way.  The abscissa
represents the total amount of extinction at 5007~\AA\ perpendicular
to the galactic plane.  Note that no matter how much extinction is
present in M101, the derived PNLF distance never varies by more than
0.1~mag from the case of no internal extinction.

Figure 5:  Images of the region around
SN~1970G\null.  The on-band image is on the left, the off-band image
is on the right.  North is up, and east is to the left; the regions displayed
are $23\parcsec 5$ on a side.  The $2\parcsec 35$ radius circles are centered 
around the estimated position of the supernova.  The supernova is 
present in the off-band, but is invisible on the on-band frame. 

Figure 6:  Images of the region around
SN~1951H\null.  The on-band image is on the left; the off-band image
is on the right. North is up, and east is to the left; the regions shown 
are $47\arcsec$ on a side.  The plus signs in the center of the images are 
at the position of the supernova, as given by Barbon, Cappellaro, \& Turatto 
(1989). The plus sign to the left and slightly above the image centers, are 
our best estimate of the supernova position from visual inspection of an 
historical image. 

Figure 7:  A plot showing the distances to galaxies with both PNLF
and Cepheid distance determinations.  Filled circles represent
direct comparisons between galaxies.  Open circles show comparisons of
different galaxies within the same cluster, and have arbitrary 1~Mpc error
bars.  Note the extremely good agreement over two orders of magnitude.  With 
the exception of M96, the scatter about the residuals is entirely consistent 
with the errors. 

\endcaptions
\vfill\eject

\bye